\documentclass[twocolumn,showpacs,amsmath,amssymb]{revtex4}

\usepackage{graphics}
\usepackage{graphicx}

\begin{document}

\title{Thermocapillary valve for droplet production and sorting}

\author{Charles N. Baroud,$^{1}$ Jean-Pierre Delville$^{2}$,
Fran\c{c}ois Gallaire$^{3}$, \& R\'egis Wunenburger$^{2}$}
\affiliation{
 $^{1}$LadHyX, Ecole Polytechnique, 91128 Palaiseau cedex, France\\
 $^{2}$CPMOH, UMR CNRS 5798, Universit\'e de Bordeaux 1, 351 Cours de
la Lib\'eration, F-33405 Talence cedex, France\\
 $^{3}$Laboratoire J.A. Dieudonn\'e, Universit\'e de Nice Sophia-Antipolis,
06108 Nice cedex, France}

\date{\today}

\begin{abstract}
Droplets are natural candidates for use as microfluidic reactors, if
active control of their formation and transport can be achieved. We show
here that localized heating from a laser can block the motion of a
water-oil interface, acting as a microfluidic valve for two-phase
flows. A theoretical model is developed to explain the forces acting on
a drop due to thermocapillary flow, predicting a scaling law which
favors miniaturization. Finally, we show how the laser forcing
can be applied to sorting drops, thus demonstrating how it may be
integrated in complex droplet microfluidic systems.
\end{abstract}

\pacs{47.61.-k, 47.55.dm}

\maketitle


Microfluidic droplets have been proposed as microreactors with the aim
to provide high performance tools for biochemistry. Individual
drops may be viewed as containing one digital bit of information and the
manipulation of a large number of slightly differing drops would allow
the testing of a large library of genes rapidly and with a small total
quantity of material~\cite{miller06}. In microchannels, drops are
produced and transported using a carrier fluid~\cite{thorsen01} and
typical channel sizes allow the manipulation of volumes in the
picoliter range. Surfactant in the carrier fluid prevents
cross-contamination of the drops through wall contact or
fusion~\cite{song03,joanicot05}. However, while the geometry of the
microchannel may be used to determine the evolution of drops and their
contents~\cite{song03, joanicot05, link04}, the implementation of real
lab-on-a-chip devices hinges on the active control of drop formation and
its evolution, which remains elusive.

In this letter, we remedy the situation by demonstrating experimentally
how a focused laser can provide precise control over droplets through
the generation of a thermocapillary flow. In doing so, we develop the
first theoretical model of a droplet subjected to localized heating,
yielding a general understanding of the forces acting on the drop and a
scaling law which favors miniaturization. A carrier fluid is still used
for the formation and transport of drops but the effects of geometry are
augmented with a local thermal gradient produced by the laser beam,
focused through a microscope objective inside the microchannel.

Indeed, moving drops with heat has been a preoccupation of fluid
mechanicians since the initial work of Young et
al~\cite{young59}. Although originally motivated by microgravity
conditions where surface effects are
dominant~\cite{young59,balasubramaniam00}, microfluidics has opened up a
new area where bulk phenomena are negligible compared to surface
effects. Recently, thermal manipulation of drops or thin films resting
on a solid substrate has received the attention of the microfluidics
community either through the embedding of electrodes in the
solid~\cite{sammarco99,darhuber03} or through optical
techniques~\cite{garnier03,sur04,kotz04}. However, the physical
mechanisms in the transmission of forces when the liquid touches a solid
wall are fundamentally different from the case of drops suspended in a
carrier fluid, away from the boundaries~\cite{lajeunesse03}. The latter
case has received little attention despite the advantages that
microchannels offer over open geometries.

Our experimental setup consists of a microchannel fabricated using soft
lithography techniques~\cite{duffy98}. Water and oil (Hexadecane + 2\%
w/w Span 80, a surfactant) are pumped into the channel at constant
flowrates, $Q_{\tt{water}}$ and $Q_{\tt{oil}}$, using glass syringes and
syringe pumps. Channel widths are in the range $100-500~\mu$m and the
height $h$ is in the range $25-50~\mu$m. Local heating is produced by a
continuous Argon-Ion laser (wavelength in vacuum
$\lambda_{Ar^{+}}=514$~nm), in the TEM$_{00}$ mode, focused inside the
channel through a $\times5$ or $\times10$ microscope objective to a beam
waist $\omega_0=5.2$ or $2.6~\mu$m, respectively. The optical approach
can be reconfigured in real-time and it allows the manipulation inside
small microchannels with no special micro-fabrication. The absorption of
the laser radiation by the aqueous phase is induced by adding 0.1\% w/w
of fluorescein in the water.


\begin{figure}
\includegraphics[width=\linewidth]{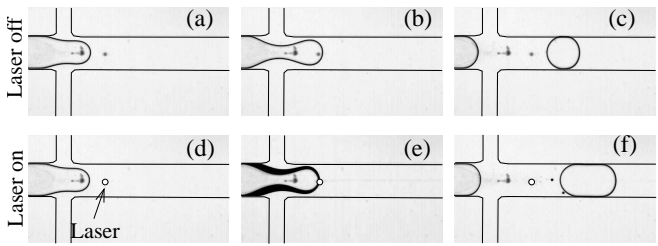}
\caption{ Microfluidic valve: In a cross-shaped microchannel, the oil
flows from the lateral channels and the water enters through the central
channel. (a)-(c) In the absence of laser forcing, drops are shed with a
typical break-off time [(b) to (c)] of 0.1~s. (d)-(f) When the laser is
applied, the interface is blocked for several seconds, producing a
larger drop. In image (e), the evolution of the neck shape is shown
through a superposition of 100 images (2~s). Exit channel width is
$200~\mu$m. Operating conditions are: $Q_{\tt{water}}=0.08~\mu$L/min,
$Q_{\tt{oil}}=0.90~\mu$L/min, beam power $P=80$~mW and beam waist
$\omega_0=5.2~\mu$m.\label{fig:valve}}
\end{figure}

A surprising effect is observed when the water-oil interface reaches the
laser spot: In the cross-shaped microchannel of Fig.~\ref{fig:valve}, we
produce water drops in oil through the hydrodynamic focusing technique
in which two oil flows pinch off water droplets at the intersection of
the channels. In the absence of the laser, drops of water are produced
in a steady fashion and are transported along with the oil down the
drain channel, as shown in Figs.~\ref{fig:valve}(a)-(c). When the laser
is illuminated, however, the oil-water interface is blocked in place as
soon as it crosses the beam. While the typical drop pinching time is
$\tau_d\sim100$~ms in the absence of the laser, we find that we can
block the interface for a time $\tau_b$ which may be of several seconds,
as shown in Figs.~\ref{fig:valve}(d)-(f) (see supporting video
1). During the time $\tau_b$, the drop shedding is completely inhibited
and the volume in the water tip increases until the viscous stresses
finally break it off. The drop thus produced is larger, since it has
been ``inflated'' by the water flow.

\begin{figure}
\centering{\includegraphics[width=8cm]{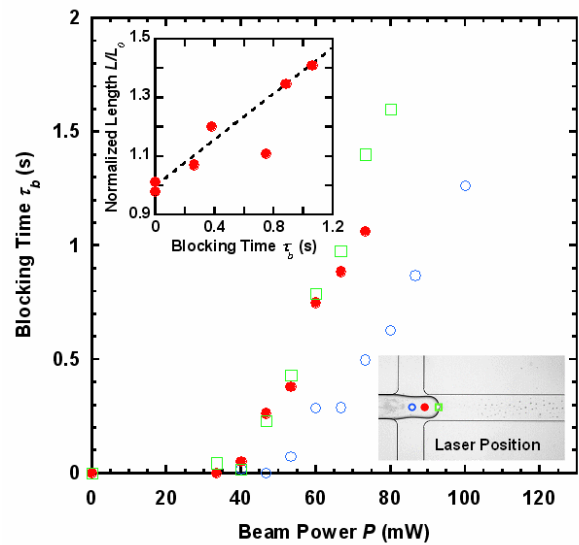}}
\caption{(color online) Dependence of the blocking time $\tau_b$ on laser power
and position (indicated in the picture) for $Q_{\tt{water}}=0.03~\mu$L/min
and $Q_{\tt{oil}}=0.1~\mu$L/min, $\omega_0=2.6~\mu$m. Inset: Rescaled droplet length $L/L_0$
vs. the blocking time (laser position $\bullet$), where $L_0$ is the
droplet length without the laser. The dashed line is a linear fit,
ignoring the outlier at $\tau_b=0.75$~s.\label{fig:Tp}}
\end{figure}

In the microchannel shown in Fig.~\ref{fig:Tp}, we measured the
variation of the blocking time $\tau_b$ with respect to laser power and
forcing position. We observe that $\tau_b$ increases approximately
linearly with the power, above an initial threshold, showing a weak
position-dependence of the laser spot. Furthermore, the inset of
Fig.~\ref{fig:Tp} shows that the droplet length $L$ varies linearly with
$\tau_b$, as expected from mass conservation at constant water flowrate
$L = L_0+\tau_b Q_{\tt{water}}/S$, $L_0$ being the droplet length
without laser and $S\simeq(125\times30)~\mu$m$^2$ the channel cross
section. The best linear fit to the data gives an effective water
flowrate $Q_{\tt{water}}=0.028~\mu$L/min, close to the nominal value
$0.03~\mu$L/min, showing that the water flowrate remains controlled even
in presence of the laser forcing. Thus, the optical forcing provides a
tunable valve which provides control over droplet timing and size.
Similar blocking is observed in a \texttt{T} geometry or if the flows
are driven at constant pressure. However, the blocking is only obtained
when the light is absorbed, here by using a dye.

\begin{figure}[ht]
\centering{\includegraphics[width=7.4cm]{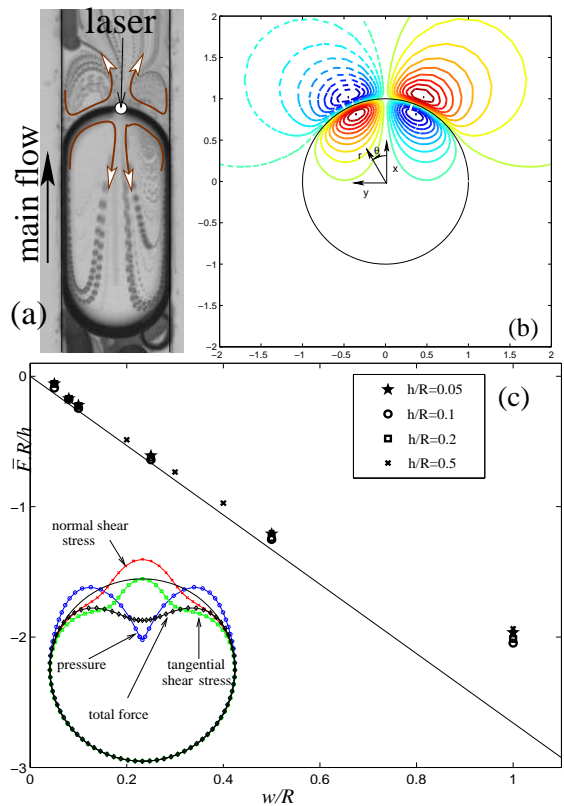}} 
\caption{(color online) (a) Overlay of 100 images from a video sequence
showing the motion of seeding particles near the hot spot. Note that the
motion along the interface is directed towards the hot spot. (b)
Streamfunction contours obtained from the depth-averaged model described
in the text. Dashed and continuous contours indicate counterclockwise
and clockwise flows, respectively. (c) Rescaled nondimensional force
$\bar{F} R/h$ plotted as a function of $w/R$ for various aspect ratios
$h/R$ for $\bar{\mu}=3/4$. The straight line corresponds to the
dimensional scaling derived in the text. The inset shows the x-component
of the distribution along the azimuthal direction of the pressure,
normal and tangential shear stresses, where the solid circle is the
reference zero. Their sum yields the total force. Channel width in part
(a) is 140$~\mu$m. $h/R=0.2$, $w/R=0.5$ for parts (b) and
(c)inset.\label{fig:model}} 
\end{figure}

We visualize the convection rolls produced by the heating by placing
tracer particles in both fluids, as shown in Fig.~\ref{fig:model}(a) for
a drop that is blocked in a straight channel. For pure liquids, the
direction of Marangoni flow along the interface is directed from the hot
(low surface tension) to the cold (high surface tension)
regions. However, the flows in our experiments point towards the laser
along the interface, indicating an increase of surface tension with
temperature. This is consistent with previous studies that have shown a
linear increase of surface tension with temperature in the presence of
surfactants~\cite{berge94,sloutskin05,surf_conc}.

%
%

One important constraint for practical applications is the amplitude of
the temperature rise. Since the materials used in this study have
similar thermal properties ({thermal diffusivity} $D_{\tt{th}} \sim
10^{-7}$~m$^2$s$^{-1}$, {thermal conductivity} $\Lambda_{\tt{th}}\sim
0.5$~{Wm$^{-1}$K$^{-1}$}), we estimate the maximum temperature in the
flow by modeling the heating produced by a laser absorbed in a single
fluid phase~\cite{gordon65}, assuming thermal diffusion as the only
energy transport mechanism. Considering the measured optical absorption
of our water/dye solution $\alpha_{\tt{th}}=117.9$~m$^{-1}$, and
assuming that the temperature $100~\mu$m away is fixed by the flowing
oil at room temperature, we find $\Delta T\simeq12$~K for the
temperature rise at the laser focus for a beam power P = 100 mW. The
temperature gradient is steep near the focus, with the temperature
dropping to 5~K at $20~\mu$m from the beam spot. However, note that
given the typical flow velocity ($U\sim1$~mm/s) and the characteristic
length scale over which thermal diffusion occurs ($L=100~\mu$m), the
thermal Peclet number $Pe=UL/D_{th}$ is comparable to unity. Thus, our
calculation overestimates the actual overheating.

%
%

The force generated by the convective flow on a droplet is investigated
through the depth-averaged Stokes equations, since our channels have a
large width/height aspect ratio~\cite{boos97}. The detailed modeling
will be discussed in a subsequent publication; here we limit ourselves
to the main features: a circular drop of radius $R$ is considered in an
infinite domain and the flow due to the Marangoni stresses is
evaluated. Assuming a parabolic profile in the small dimension ($h$) and
introducing a streamfunction for the mean velocities in the plane of the
channel, the depth averaged equations, valid in each fluid, are

\begin{equation}
\left(\frac{1}{r} \frac{\partial }{\partial
r}r\frac{\partial}{\partial r}
+\frac{1}{r^2}\frac{\partial^2}{\partial\theta^2}\right)
\left(\frac{1}{r} \frac{\partial }{\partial
r}r\frac{\partial}{\partial r}
+\frac{1}{r^2}\frac{\partial^2}{\partial\theta^2}-\frac{12}{h^2}\right)
\psi=0,
\end{equation}
where the depth-averaged velocities may be retrieved from
$u_\theta=-{\partial \psi}/{\partial r}$ and $u_r={1}/{r}({\partial
\psi}/{\partial \theta})$. The kinematic boundary conditions at the drop
interface ($r=R$) are zero normal velocity
\noindent and the continuity of the tangential velocity. The normal
dynamic boundary condition is not imposed since the drop is assumed to
remain circular, which is consistent with our experimental
observations, Fig.~3a. Finally, the tangential dynamic boundary
condition, which accounts for the optically-induced Marangoni stress, is

\begin{equation}
\mu_1 r \frac{\partial }{ \partial
r}\left(\frac{u_\theta^1}{r}\right) -\mu_2 r  \frac{\partial }{
\partial r}\left(\frac{u_\theta^2}{r}\right)= -\frac{\gamma'}{r} \frac{d T}{d
\theta},
\label{eq:marangoni}
\end{equation}
where $\mu_{1,2}$ are the dynamic viscosities and $u_\theta^{1,2}$ are
the velocities in the drop and the carrier fluid,
respectively. $\gamma'=\partial \gamma/\partial T$ is the surface
tension to temperature gradient, which is positive in our case.

For simplicity, we approximate the steady state temperature distribution
using a Gaussian form $T(x,y)=\Delta T \exp[-((x-R)^2+y^2)/{w^2}]$,
where $\Delta T$ is the maximum temperature difference between the hot
spot and the far field and $w$ corresponds to the size of the diffused
hot spot, which is significantly larger than
$\omega_0$~\cite{gordon65}. The equations are nondimensionalized using
$\Delta T$ as temperature scale, $R$ as length scale, $R\gamma' \Delta
T$ as force scale and $\frac{R(\mu_1+\mu_2)}{\gamma' \Delta T}$ as time
scale, the remaining nondimensional groups being the aspect ratio $h/R$,
the nondimensional hot spot size $w/R$ and the viscosity ratio
$\bar{\mu}=\mu_2/(\mu_1+\mu_2)$.

A typical predicted flow field solving the above numerical formulation
is shown in Fig.~\ref{fig:model}(b), in which the four recirculation
regions are clearly visible. The velocity gradients display a separation
of scales in the normal and tangential directions, as observed from the
distance between the streamlines in the two directions. Indeed, it may
be verified that the velocities vary over a typical length scale $h/R$
in the normal direction, while the tangential length scale is given by
$w/R$.

Along with this flow field, we compute the pressure field, as well as
the normal ($\bar{\sigma}_{\bar{r}\bar{r}}=2\bar{\mu}\frac{\partial
\bar{u}_{\bar{r}}}{\partial \bar{r}}$) and tangential
$\left(\bar{\sigma}_{\bar{r}\theta}=\bar{\mu}\left(\frac{1}{\bar{r}}
\frac{\partial \bar{u}_{\bar{r}}}{\partial \theta} + \frac{\partial
\bar{u}_\theta}{\partial \bar{r}} - \frac{
\bar{u}_\theta}{\bar{r}}\right)\right)$ viscous shear stresses in the
external flow. Their projections on the $x$ axis, shown in the inset of
Fig.~\ref{fig:model}(c), are then summed and integrated along $\theta$
to yield the total dimensionless force ($\bar{F}$) on the drop. Note
that the global $x$ component of the force is negative and therefore
opposes the transport of the drop by the external flow. The $y$
component vanishes by symmetry and the integral of the wall friction may
be shown to be zero since the drop is stationary. Numerically computed
values of $\bar{F}R/h$ are shown by the isolated points in
Fig.~\ref{fig:model}(c) as a function of $w/R$, for different values of
the aspect ratio $h/R$. The points all collapse on a single master
curve, displaying a nondimensional scaling law $\bar{F} \propto wh/R^2$,
for small $w/R$.


The dimensional form of the force can be obtained, for small $h/R$ and
$w/R$, by considering the three contributions separately and noting that
the velocity scale in this problem is imposed by the Marangoni
stress. Using the separation of scales along the azimuthal and radial
directions, Eq.~\ref{eq:marangoni} becomes
$\left(\mu_1+\mu_2\right)\frac{U}{h}\sim\frac{\Delta T \gamma'}{R}
\frac{R}{w}$, where the '$\sim$' is understood as an
order-of-magnitude scaling. This yields the characteristic tangential
velocity scale

\begin{equation} 
U \sim \frac{\Delta T \gamma'}{\mu_1+\mu_2}\frac{h}{w}. 
\end{equation} 

\noindent The force due to the tangential viscous shear is then obtained
by multiplying $\sigma_{r\theta}\sim\mu_2U/h$ by $\sin\theta\simeq w/R$
and integrating on the portion $w\times h$ of the interface,

\begin{equation} 
F_t\sim \mu_2 \frac{U}{h} \frac {w}{R} w h =
\frac{\mu_2}{\mu_1+\mu_2} \Delta T \gamma' \frac{hw}{R}.  
\label{eq:scaling}
\end{equation}

\noindent The force due to the normal viscous shear can be shown to
scale like $F_n\sim\frac{h}{R}F_t$ and is therefore negligible. The
pressure force, on the other hand, derives from a balance between the
pressure gradient and the radial second derivative of velocity. In the
present circular geometry, similar scaling arguments yield a law for the
contribution of the pressure force $F_p$, which follows the same scaling
as $F_t$, resulting in the same scaling law for the total force $F$. 
A rigorous derivation (to be published elsewhere)
yields the final form of the force including the prefactor:

\begin{equation}
F=-2\sqrt{\pi}\frac{\mu_2}{\mu_1+\mu_2} \Delta T \gamma'\frac{hw}{R}.
\end{equation}
 This expression is represented (once non-dimensionalized)
by the straight line on Fig.~\ref{fig:model}(c) and agrees very well
with the numerically computed values.

The physical value of the force for a typical experiment is estimated by
taking $\mu_1=10^{-3}$~Nm$^{-2}$s (water), $\mu_2=3\mu_1$ (hexadecane),
and extracting $\gamma'\sim1$~mNm$^{-1}$K$^{-1}$ from
Ref.~\cite{sloutskin05}. This yields a force on the order of 0.1~$\mu$N,
which is of the same order as the drag force on a drop in a large aspect
ratio channel~\cite{nadim96}, thus confirming that thermocapillary
forcing can indeed account for the blocking. Note that the force we
calculate is several orders of magnitude larger those generated from
electric fields~\cite{ahn06a} or optical tweezers~\cite{grier03}.

%
%

\begin{figure}
\centering{\rotatebox{-90}{\includegraphics[width=5.6cm]{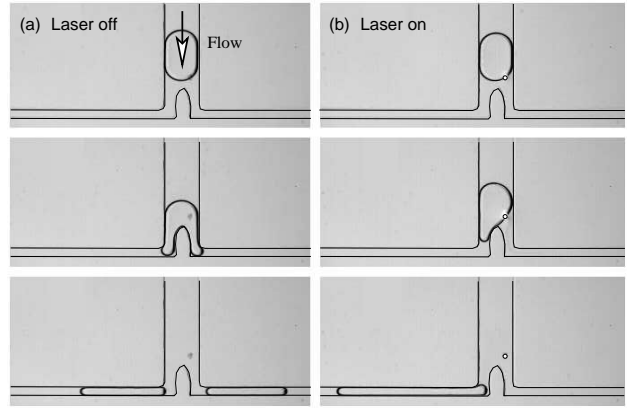}}}
\caption{Sorting drops: (a) Without laser forcing, a drop at a bifurcation
divides into approximately equal daughter droplets. (b) {When the
laser forcing is applied, the drop advance in the right hand channel is
blocked so the whole drop is diverted into the left channel.} Main
channel width is $200~\mu$m and the operating conditions are
$Q_{\tt{water}}=0.02~\mu$L/min, $Q_{\tt{oil}}=0.2~\mu$L/min, and
$\omega_0=5.2~\mu$m.
\label{fig:sorting} }
\end{figure}

This blocking force may be applied at different locations in a
microchannel by displacing the laser spot. In particular, we demonstrate
the sorting of drops, a fundamental operation in the implementation of
lab-on-a-chip devices. Drops are formed, as above, in a cross-junction
and arrive at a symmetric bifurcation, carried by the continuous
phase. In the absence of laser forcing, the drops arriving at the
bifurcation divide into two equal parts~\cite{link04},
Fig.~\ref{fig:sorting}(a). When the laser is applied, the water-oil
interface is asymmetrically blocked on the right hand side while the
left hand side continues to flow, Fig.~\ref{fig:sorting}(b). Above a
critical laser power (approximately 100~mW for the present
configuration), the drop is blocked long enough that it is completely
diverted through the left hand channel (see supporting video 2). Drops
may therefore be sorted by accordingly selecting the laser position.

%
%

In summary, we have experimentally and theoretically demonstrated the
efficiency of laser-driven blocking of water-in-oil drops. The
theoretical treatment brings out two length scales, $h/R$ and
$w/R$. While $h$ and $w$ can be thought of as determining the typical
scales for velocity variations in the radial and azimuthal directions,
$R$ enters the force scaling as a local radius of curvature rather than
the actual size of the drop. It is therefore not surprising that the
blocking force should {\it increase} as $R$ decreases. On the other
hand, the drag force due to the external flow scales as
$R^2$~\cite{nadim96}, implying that the laser power necessary to
counterbalance the drag quickly decreases with the drop size. This,
along with the rapidity of viscous and thermal diffusion while thermal
inertia is reduced, all lead to laws favorable to miniaturization. The
generality of the process provides a practical new way for acting on
individual droplets, at any location, while working inside the robust
environment of the microchannel.

We acknowledge help from Julien Buchoux, David Dulin and Emilie
Verneuil. This work was partially funded by the CNRS {\it PIR}
``Microfluidique et Microsyst\`emes Fluidiques'', the Conseil
R\'egional d'Aquitaine, and the {\it convention X-DGA}.


\begin{thebibliography}{22}
\bibitem[{sur()}]{surf_conc}
\bibinfo{note}{A reduction the surfactant concentration reduced the efficiency
  of the blocking; a detailed exploration of surfactant effects will be
  addressed in a future study.}
\end{thebibliography}
\end{document}